\pdfoutput=1
\documentclass[11pt,twoside,a4paper,cmspaper,final,collab]{cms-tdr}
\def\svnVersion{29318:29951}\def\cmsCernNo{2010-067}\def\cmsCernDate{\today}\def\cmsMessage{Submitted to the Journal of High Energy Physics}

\begin{document}\cmsNoteHeader{EXO-10-011}
\hyphenation{env-iron-men-tal}
\hyphenation{had-ron-i-za-tion}
\hyphenation{cal-or-i-me-ter}
\hyphenation{de-vices}
\RCS$Revision: 29945 $
\RCS$HeadURL: svn+ssh://alverson@svn.cern.ch/reps/tdr2/papers/EXO-10-011/trunk/EXO-10-011.tex $
\RCS$Id: EXO-10-011.tex 29945 2011-01-09 12:06:07Z gbruno $
%
%
%

\providecommand {\etal}{\mbox{et al.}\xspace} 
\providecommand {\ie}{\mbox{i.e.}\xspace}     
\providecommand {\eg}{\mbox{e.g.}\xspace}     
\providecommand {\etc}{\mbox{etc.}\xspace}     
\providecommand {\vs}{\mbox{\sl vs.}\xspace}      
\providecommand {\mdash}{\ensuremath{\mathrm{-}}} 

\providecommand {\Lone}{Level-1\xspace} 
\providecommand {\Ltwo}{Level-2\xspace}
\providecommand {\Lthree}{Level-3\xspace}

\providecommand{\ACERMC} {\textsc{AcerMC}\xspace}
\providecommand{\ALPGEN} {{\textsc{alpgen}}\xspace}
\providecommand{\CHARYBDIS} {{\textsc{charybdis}}\xspace}
\providecommand{\CMKIN} {\textsc{cmkin}\xspace}
\providecommand{\CMSIM} {{\textsc{cmsim}}\xspace}
\providecommand{\CMSSW} {{\textsc{cmssw}}\xspace}
\providecommand{\COBRA} {{\textsc{cobra}}\xspace}
\providecommand{\COCOA} {{\textsc{cocoa}}\xspace}
\providecommand{\COMPHEP} {\textsc{CompHEP}\xspace}
\providecommand{\EVTGEN} {{\textsc{evtgen}}\xspace}
\providecommand{\FAMOS} {{\textsc{famos}}\xspace}
\providecommand{\GARCON} {\textsc{garcon}\xspace}
\providecommand{\GARFIELD} {{\textsc{garfield}}\xspace}
\providecommand{\GEANE} {{\textsc{geane}}\xspace}
\providecommand{\GEANTfour} {{\textsc{geant4}}\xspace}
\providecommand{\GEANTthree} {{\textsc{geant3}}\xspace}
\providecommand{\GEANT} {{\textsc{geant}}\xspace}
\providecommand{\HDECAY} {\textsc{hdecay}\xspace}
\providecommand{\HERWIG} {{\textsc{herwig}}\xspace}
\providecommand{\HIGLU} {{\textsc{higlu}}\xspace}
\providecommand{\HIJING} {{\textsc{hijing}}\xspace}
\providecommand{\IGUANA} {\textsc{iguana}\xspace}
\providecommand{\ISAJET} {{\textsc{isajet}}\xspace}
\providecommand{\ISAPYTHIA} {{\textsc{isapythia}}\xspace}
\providecommand{\ISASUGRA} {{\textsc{isasugra}}\xspace}
\providecommand{\ISASUSY} {{\textsc{isasusy}}\xspace}
\providecommand{\ISAWIG} {{\textsc{isawig}}\xspace}
\providecommand{\MADGRAPH} {\textsc{MadGraph}\xspace}
\providecommand{\MCATNLO} {\textsc{mc@nlo}\xspace}
\providecommand{\MCFM} {\textsc{mcfm}\xspace}
\providecommand{\MILLEPEDE} {{\textsc{millepede}}\xspace}
\providecommand{\ORCA} {{\textsc{orca}}\xspace}
\providecommand{\OSCAR} {{\textsc{oscar}}\xspace}
\providecommand{\PHOTOS} {\textsc{photos}\xspace}
\providecommand{\PROSPINO} {\textsc{prospino}\xspace}
\providecommand{\PYTHIA} {{\textsc{pythia}}\xspace}
\providecommand{\SHERPA} {{\textsc{sherpa}}\xspace}
\providecommand{\TAUOLA} {\textsc{tauola}\xspace}
\providecommand{\TOPREX} {\textsc{TopReX}\xspace}
\providecommand{\XDAQ} {{\textsc{xdaq}}\xspace}

\providecommand {\DZERO}{D\O\xspace}     


\providecommand{\de}{\ensuremath{^\circ}}
\providecommand{\ten}[1]{\ensuremath{\times \text{10}^\text{#1}}}
\providecommand{\unit}[1]{\ensuremath{\text{\,#1}}\xspace}
\providecommand{\mum}{\ensuremath{\,\mu\text{m}}\xspace}
\providecommand{\micron}{\ensuremath{\,\mu\text{m}}\xspace}
\providecommand{\cm}{\ensuremath{\,\text{cm}}\xspace}
\providecommand{\mm}{\ensuremath{\,\text{mm}}\xspace}
\providecommand{\mus}{\ensuremath{\,\mu\text{s}}\xspace}
\providecommand{\keV}{\ensuremath{\,\text{ke\hspace{-.08em}V}}\xspace}
\providecommand{\MeV}{\ensuremath{\,\text{Me\hspace{-.08em}V}}\xspace}
\providecommand{\GeV}{\ensuremath{\,\text{Ge\hspace{-.08em}V}}\xspace}
\providecommand{\gev}{\GeV}
\providecommand{\TeV}{\ensuremath{\,\text{Te\hspace{-.08em}V}}\xspace}
\providecommand{\PeV}{\ensuremath{\,\text{Pe\hspace{-.08em}V}}\xspace}
\providecommand{\keVc}{\ensuremath{{\,\text{ke\hspace{-.08em}V\hspace{-0.16em}/\hspace{-0.08em}}c}}\xspace}
\providecommand{\MeVc}{\ensuremath{{\,\text{Me\hspace{-.08em}V\hspace{-0.16em}/\hspace{-0.08em}}c}}\xspace}
\providecommand{\GeVc}{\ensuremath{{\,\text{Ge\hspace{-.08em}V\hspace{-0.16em}/\hspace{-0.08em}}c}}\xspace}
\providecommand{\TeVc}{\ensuremath{{\,\text{Te\hspace{-.08em}V\hspace{-0.16em}/\hspace{-0.08em}}c}}\xspace}
\providecommand{\keVcc}{\ensuremath{{\,\text{ke\hspace{-.08em}V\hspace{-0.16em}/\hspace{-0.08em}}c^\text{2}}}\xspace}
\providecommand{\MeVcc}{\ensuremath{{\,\text{Me\hspace{-.08em}V\hspace{-0.16em}/\hspace{-0.08em}}c^\text{2}}}\xspace}
\providecommand{\GeVcc}{\ensuremath{{\,\text{Ge\hspace{-.08em}V\hspace{-0.16em}/\hspace{-0.08em}}c^\text{2}}}\xspace}
\providecommand{\TeVcc}{\ensuremath{{\,\text{Te\hspace{-.08em}V\hspace{-0.16em}/\hspace{-0.08em}}c^\text{2}}}\xspace}

\providecommand{\pbinv} {\mbox{\ensuremath{\,\text{pb}^\text{$-$1}}}\xspace}
\providecommand{\fbinv} {\mbox{\ensuremath{\,\text{fb}^\text{$-$1}}}\xspace}
\providecommand{\nbinv} {\mbox{\ensuremath{\,\text{nb}^\text{$-$1}}}\xspace}
\providecommand{\percms}{\ensuremath{\,\text{cm}^\text{$-$2}\,\text{s}^\text{$-$1}}\xspace}
\providecommand{\lumi}{\ensuremath{\mathcal{L}}\xspace}
\providecommand{\Lumi}{\ensuremath{\mathcal{L}}\xspace}
%
\providecommand{\LvLow}  {\ensuremath{\mathcal{L}=\text{10}^\text{32}\,\text{cm}^\text{$-$2}\,\text{s}^\text{$-$1}}\xspace}
\providecommand{\LLow}   {\ensuremath{\mathcal{L}=\text{10}^\text{33}\,\text{cm}^\text{$-$2}\,\text{s}^\text{$-$1}}\xspace}
\providecommand{\lowlumi}{\ensuremath{\mathcal{L}=\text{2}\times \text{10}^\text{33}\,\text{cm}^\text{$-$2}\,\text{s}^\text{$-$1}}\xspace}
\providecommand{\LMed}   {\ensuremath{\mathcal{L}=\text{2}\times \text{10}^\text{33}\,\text{cm}^\text{$-$2}\,\text{s}^\text{$-$1}}\xspace}
\providecommand{\LHigh}  {\ensuremath{\mathcal{L}=\text{10}^\text{34}\,\text{cm}^\text{$-$2}\,\text{s}^\text{$-$1}}\xspace}
\providecommand{\hilumi} {\ensuremath{\mathcal{L}=\text{10}^\text{34}\,\text{cm}^\text{$-$2}\,\text{s}^\text{$-$1}}\xspace}


\providecommand{\PT}{\ensuremath{p_{\mathrm{T}}}\xspace}
\providecommand{\pt}{\ensuremath{p_{\mathrm{T}}}\xspace}
\providecommand{\ET}{\ensuremath{E_{\mathrm{T}}}\xspace}
\providecommand{\HT}{\ensuremath{H_{\mathrm{T}}}\xspace}
\providecommand{\et}{\ensuremath{E_{\mathrm{T}}}\xspace}
\providecommand{\Em}{\ensuremath{E\hspace{-0.6em}/}\xspace}
\providecommand{\Pm}{\ensuremath{p\hspace{-0.5em}/}\xspace}
\providecommand{\PTm}{\ensuremath{{p}_\mathrm{T}\hspace{-1.02em}/}\xspace}
\providecommand{\PTslash}{\ensuremath{{p}_\mathrm{T}\hspace{-1.02em}/}\xspace}
\providecommand{\ETm}{\ensuremath{E_{\mathrm{T}}^{\text{miss}}}\xspace}
\providecommand{\MET}{\ETm}
\providecommand{\ETmiss}{\ETm}
\providecommand{\ETslash}{\ensuremath{E_{\mathrm{T}}\hspace{-1.1em}/}\xspace}
\providecommand{\VEtmiss}{\ensuremath{{\vec E}_{\mathrm{T}}^{\text{miss}}}\xspace}

\providecommand{\dd}[2]{\ensuremath{\frac{\mathrm{d} #1}{\mathrm{d} #2}}}
\providecommand{\ddinline}[2]{\ensuremath{\mathrm{d} #1/\mathrm{d} #2}}

\ifthenelse{\boolean{cms@italic}}{\newcommand{\cmsSymbolFace}{\relax}}{\newcommand{\cmsSymbolFace}{\mathrm}}

\providecommand{\zp}{\ensuremath{\cmsSymbolFace{Z}^\prime}\xspace}
\providecommand{\JPsi}{\ensuremath{\cmsSymbolFace{J}\hspace{-.08em}/\hspace{-.14em}\psi}\xspace}
\providecommand{\Z}{\ensuremath{\cmsSymbolFace{Z}}\xspace}
\providecommand{\ttbar}{\ensuremath{\cmsSymbolFace{t}\overline{\cmsSymbolFace{t}}}\xspace}

\newcommand{\cPgn}{\ensuremath{\nu}}
\newcommand{\cPJgy}{\JPsi}
\newcommand{\cPZ}{\Z}
\newcommand{\cPZpr}{\zp}


\providecommand{\AFB}{\ensuremath{A_\text{FB}}\xspace}
\providecommand{\wangle}{\ensuremath{\sin^{2}\theta_{\text{eff}}^\text{lept}(M^2_\Z)}\xspace}
\providecommand{\stat}{\ensuremath{\,\text{(stat.)}}\xspace}
\providecommand{\syst}{\ensuremath{\,\text{(syst.)}}\xspace}
\providecommand{\kt}{\ensuremath{k_{\mathrm{T}}}\xspace}

\providecommand{\BC}{\ensuremath{\mathrm{B_{c}}}\xspace}
\providecommand{\bbarc}{\ensuremath{\mathrm{\overline{b}c}}\xspace}
\providecommand{\bbbar}{\ensuremath{\mathrm{b\overline{b}}}\xspace}
\providecommand{\ccbar}{\ensuremath{\mathrm{c\overline{c}}}\xspace}
\providecommand{\bspsiphi}{\ensuremath{\mathrm{B_s} \to \JPsi\, \phi}\xspace}
\providecommand{\EE}{\ensuremath{\mathrm{e^+e^-}}\xspace}
\providecommand{\MM}{\ensuremath{\mu^+\mu^-}\xspace}
\providecommand{\TT}{\ensuremath{\tau^+\tau^-}\xspace}

\providecommand{\HGG}{\ensuremath{\mathrm{H}\to\gamma\gamma}}
\providecommand{\GAMJET}{\ensuremath{\gamma + \text{jet}}}
\providecommand{\PPTOJETS}{\ensuremath{\mathrm{pp}\to\text{jets}}}
\providecommand{\PPTOGG}{\ensuremath{\mathrm{pp}\to\gamma\gamma}}
\providecommand{\PPTOGAMJET}{\ensuremath{\mathrm{pp}\to\gamma + \mathrm{jet}}}
\providecommand{\MH}{\ensuremath{M_{\mathrm{H}}}}
\providecommand{\RNINE}{\ensuremath{R_\mathrm{9}}}
\providecommand{\DR}{\ensuremath{\Delta R}}

%

\providecommand{\ga}{\ensuremath{\gtrsim}}
\providecommand{\la}{\ensuremath{\lesssim}}
\providecommand{\swsq}{\ensuremath{\sin^2\theta_\cmsSymbolFace{W}}\xspace}
\providecommand{\cwsq}{\ensuremath{\cos^2\theta_\cmsSymbolFace{W}}\xspace}
\providecommand{\tanb}{\ensuremath{\tan\beta}\xspace}
\providecommand{\tanbsq}{\ensuremath{\tan^{2}\beta}\xspace}
\providecommand{\sidb}{\ensuremath{\sin 2\beta}\xspace}
\providecommand{\alpS}{\ensuremath{\alpha_S}\xspace}
\providecommand{\alpt}{\ensuremath{\tilde{\alpha}}\xspace}

\providecommand{\QL}{\ensuremath{\cmsSymbolFace{Q}_\cmsSymbolFace{L}}\xspace}
\providecommand{\sQ}{\ensuremath{\tilde{\cmsSymbolFace{Q}}}\xspace}
\providecommand{\sQL}{\ensuremath{\tilde{\cmsSymbolFace{Q}}_\cmsSymbolFace{L}}\xspace}
\providecommand{\ULC}{\ensuremath{\cmsSymbolFace{U}_\cmsSymbolFace{L}^\cmsSymbolFace{C}}\xspace}
\providecommand{\sUC}{\ensuremath{\tilde{\cmsSymbolFace{U}}^\cmsSymbolFace{C}}\xspace}
\providecommand{\sULC}{\ensuremath{\tilde{\cmsSymbolFace{U}}_\cmsSymbolFace{L}^\cmsSymbolFace{C}}\xspace}
\providecommand{\DLC}{\ensuremath{\cmsSymbolFace{D}_\cmsSymbolFace{L}^\cmsSymbolFace{C}}\xspace}
\providecommand{\sDC}{\ensuremath{\tilde{\cmsSymbolFace{D}}^\cmsSymbolFace{C}}\xspace}
\providecommand{\sDLC}{\ensuremath{\tilde{\cmsSymbolFace{D}}_\cmsSymbolFace{L}^\cmsSymbolFace{C}}\xspace}
\providecommand{\LL}{\ensuremath{\cmsSymbolFace{L}_\cmsSymbolFace{L}}\xspace}
\providecommand{\sL}{\ensuremath{\tilde{\cmsSymbolFace{L}}}\xspace}
\providecommand{\sLL}{\ensuremath{\tilde{\cmsSymbolFace{L}}_\cmsSymbolFace{L}}\xspace}
\providecommand{\ELC}{\ensuremath{\cmsSymbolFace{E}_\cmsSymbolFace{L}^\cmsSymbolFace{C}}\xspace}
\providecommand{\sEC}{\ensuremath{\tilde{\cmsSymbolFace{E}}^\cmsSymbolFace{C}}\xspace}
\providecommand{\sELC}{\ensuremath{\tilde{\cmsSymbolFace{E}}_\cmsSymbolFace{L}^\cmsSymbolFace{C}}\xspace}
\providecommand{\sEL}{\ensuremath{\tilde{\cmsSymbolFace{E}}_\cmsSymbolFace{L}}\xspace}
\providecommand{\sER}{\ensuremath{\tilde{\cmsSymbolFace{E}}_\cmsSymbolFace{R}}\xspace}
\providecommand{\sFer}{\ensuremath{\tilde{\cmsSymbolFace{f}}}\xspace}
\providecommand{\sQua}{\ensuremath{\tilde{\cmsSymbolFace{q}}}\xspace}
\providecommand{\sUp}{\ensuremath{\tilde{\cmsSymbolFace{u}}}\xspace}
\providecommand{\suL}{\ensuremath{\tilde{\cmsSymbolFace{u}}_\cmsSymbolFace{L}}\xspace}
\providecommand{\suR}{\ensuremath{\tilde{\cmsSymbolFace{u}}_\cmsSymbolFace{R}}\xspace}
\providecommand{\sDw}{\ensuremath{\tilde{\cmsSymbolFace{d}}}\xspace}
\providecommand{\sdL}{\ensuremath{\tilde{\cmsSymbolFace{d}}_\cmsSymbolFace{L}}\xspace}
\providecommand{\sdR}{\ensuremath{\tilde{\cmsSymbolFace{d}}_\cmsSymbolFace{R}}\xspace}
\providecommand{\sTop}{\ensuremath{\tilde{\cmsSymbolFace{t}}}\xspace}
\providecommand{\stL}{\ensuremath{\tilde{\cmsSymbolFace{t}}_\cmsSymbolFace{L}}\xspace}
\providecommand{\stR}{\ensuremath{\tilde{\cmsSymbolFace{t}}_\cmsSymbolFace{R}}\xspace}
\providecommand{\stone}{\ensuremath{\tilde{\cmsSymbolFace{t}}_1}\xspace}
\providecommand{\sttwo}{\ensuremath{\tilde{\cmsSymbolFace{t}}_2}\xspace}
\providecommand{\sBot}{\ensuremath{\tilde{\cmsSymbolFace{b}}}\xspace}
\providecommand{\sbL}{\ensuremath{\tilde{\cmsSymbolFace{b}}_\cmsSymbolFace{L}}\xspace}
\providecommand{\sbR}{\ensuremath{\tilde{\cmsSymbolFace{b}}_\cmsSymbolFace{R}}\xspace}
\providecommand{\sbone}{\ensuremath{\tilde{\cmsSymbolFace{b}}_1}\xspace}
\providecommand{\sbtwo}{\ensuremath{\tilde{\cmsSymbolFace{b}}_2}\xspace}
\providecommand{\sLep}{\ensuremath{\tilde{\cmsSymbolFace{l}}}\xspace}
\providecommand{\sLepC}{\ensuremath{\tilde{\cmsSymbolFace{l}}^\cmsSymbolFace{C}}\xspace}
\providecommand{\sEl}{\ensuremath{\tilde{\cmsSymbolFace{e}}}\xspace}
\providecommand{\sElC}{\ensuremath{\tilde{\cmsSymbolFace{e}}^\cmsSymbolFace{C}}\xspace}
\providecommand{\seL}{\ensuremath{\tilde{\cmsSymbolFace{e}}_\cmsSymbolFace{L}}\xspace}
\providecommand{\seR}{\ensuremath{\tilde{\cmsSymbolFace{e}}_\cmsSymbolFace{R}}\xspace}
\providecommand{\snL}{\ensuremath{\tilde{\nu}_L}\xspace}
\providecommand{\sMu}{\ensuremath{\tilde{\mu}}\xspace}
\providecommand{\sNu}{\ensuremath{\tilde{\nu}}\xspace}
\providecommand{\sTau}{\ensuremath{\tilde{\tau}}\xspace}
\providecommand{\Glu}{\ensuremath{\cmsSymbolFace{g}}\xspace}
\providecommand{\sGlu}{\ensuremath{\tilde{\cmsSymbolFace{g}}}\xspace}
\providecommand{\Wpm}{\ensuremath{\cmsSymbolFace{W}^{\pm}}\xspace}
\providecommand{\sWpm}{\ensuremath{\tilde{\cmsSymbolFace{W}}^{\pm}}\xspace}
\providecommand{\Wz}{\ensuremath{\cmsSymbolFace{W}^{0}}\xspace}
\providecommand{\sWz}{\ensuremath{\tilde{\cmsSymbolFace{W}}^{0}}\xspace}
\providecommand{\sWino}{\ensuremath{\tilde{\cmsSymbolFace{W}}}\xspace}
\providecommand{\Bz}{\ensuremath{\cmsSymbolFace{B}^{0}}\xspace}
\providecommand{\sBz}{\ensuremath{\tilde{\cmsSymbolFace{B}}^{0}}\xspace}
\providecommand{\sBino}{\ensuremath{\tilde{\cmsSymbolFace{B}}}\xspace}
\providecommand{\Zz}{\ensuremath{\cmsSymbolFace{Z}^{0}}\xspace}
\providecommand{\sZino}{\ensuremath{\tilde{\cmsSymbolFace{Z}}^{0}}\xspace}
\providecommand{\sGam}{\ensuremath{\tilde{\gamma}}\xspace}
\providecommand{\chiz}{\ensuremath{\tilde{\chi}^{0}}\xspace}
\providecommand{\chip}{\ensuremath{\tilde{\chi}^{+}}\xspace}
\providecommand{\chim}{\ensuremath{\tilde{\chi}^{-}}\xspace}
\providecommand{\chipm}{\ensuremath{\tilde{\chi}^{\pm}}\xspace}
\providecommand{\Hone}{\ensuremath{\cmsSymbolFace{H}_\cmsSymbolFace{d}}\xspace}
\providecommand{\sHone}{\ensuremath{\tilde{\cmsSymbolFace{H}}_\cmsSymbolFace{d}}\xspace}
\providecommand{\Htwo}{\ensuremath{\cmsSymbolFace{H}_\cmsSymbolFace{u}}\xspace}
\providecommand{\sHtwo}{\ensuremath{\tilde{\cmsSymbolFace{H}}_\cmsSymbolFace{u}}\xspace}
\providecommand{\sHig}{\ensuremath{\tilde{\cmsSymbolFace{H}}}\xspace}
\providecommand{\sHa}{\ensuremath{\tilde{\cmsSymbolFace{H}}_\cmsSymbolFace{a}}\xspace}
\providecommand{\sHb}{\ensuremath{\tilde{\cmsSymbolFace{H}}_\cmsSymbolFace{b}}\xspace}
\providecommand{\sHpm}{\ensuremath{\tilde{\cmsSymbolFace{H}}^{\pm}}\xspace}
\providecommand{\hz}{\ensuremath{\cmsSymbolFace{h}^{0}}\xspace}
\providecommand{\Hz}{\ensuremath{\cmsSymbolFace{H}^{0}}\xspace}
\providecommand{\Az}{\ensuremath{\cmsSymbolFace{A}^{0}}\xspace}
\providecommand{\Hpm}{\ensuremath{\cmsSymbolFace{H}^{\pm}}\xspace}
\providecommand{\sGra}{\ensuremath{\tilde{\cmsSymbolFace{G}}}\xspace}
\providecommand{\mtil}{\ensuremath{\tilde{m}}\xspace}
\providecommand{\rpv}{\ensuremath{\rlap{\kern.2em/}R}\xspace}
\providecommand{\LLE}{\ensuremath{LL\bar{E}}\xspace}
\providecommand{\LQD}{\ensuremath{LQ\bar{D}}\xspace}
\providecommand{\UDD}{\ensuremath{\overline{UDD}}\xspace}
\providecommand{\Lam}{\ensuremath{\lambda}\xspace}
\providecommand{\Lamp}{\ensuremath{\lambda'}\xspace}
\providecommand{\Lampp}{\ensuremath{\lambda''}\xspace}
\providecommand{\spinbd}[2]{\ensuremath{\bar{#1}_{\dot{#2}}}\xspace}

\providecommand{\MD}{\ensuremath{{M_\mathrm{D}}}\xspace}
\providecommand{\Mpl}{\ensuremath{{M_\mathrm{Pl}}}\xspace}
\providecommand{\Rinv} {\ensuremath{{R}^{-1}}\xspace} 
\cmsNoteHeader{EXO-10-011} 
\title{Search for Heavy Stable Charged Particles in pp Collisions at $\sqrt{s}=7$ TeV}
\address[cern]{CERN}
\author[cern]{The CMS Collaboration}
\date{\today}
\newcommand{\stau}{$\tilde \tau _1 \ $}
\newcommand{\smu}{$\tilde \mu _1 \ $}
\newcommand{\sel}{$\tilde e _1 \ $}
\newcommand{\stp}{$\tilde t _1 \ $}
\newcommand{\gluino}{$\tilde g \ $}
\newcommand{\ETM}{$E_t^{miss} \ $}
\newcommand{\sto}{$\tilde{t}_1$}
\newcommand{\gsim}{\rlap{\raise -.3ex\hbox{${\scriptstyle\sim}$}}%
                   \raise .6ex\hbox{${\scriptstyle >}$}}%
\abstract{
The result of a search at the LHC for heavy stable charged
particles produced in pp collisions at $\sqrt{s} = 7$ TeV is
described. 
The data sample was collected with the CMS detector and corresponds to
an integrated luminosity of $3.1$ pb$^{-1}$. Momentum and
ionization-energy-loss measurements in the inner tracker detector  
are used to identify tracks compatible with heavy
slow-moving particles. Additionally, tracks passing muon
identification requirements are also analyzed for the same signature.  
In each case, no candidate passes the selection, with an expected 
background of less than $0.1$ events. A lower limit at the
95\% confidence level on the mass of a stable gluino is 
set at 398 GeV/$c^2$, using a conventional model of nuclear
interactions that allows charged hadrons containing this particle to reach
the muon detectors. A lower limit of 311 GeV/$c^2$
is also set for a stable gluino in a conservative scenario of complete
charge suppression, where any hadron containing this particle becomes
neutral before reaching the muon detectors.} 
\hypersetup{%
pdfauthor={CMS Collaboration},%
pdftitle={Search for Heavy Stable Charged Particles in pp collisions at sqrt(s)=7 TeV},%
pdfsubject={CMS},%
pdfkeywords={CMS, physics}}

\maketitle 
\section{Introduction}
Heavy stable (or quasi-stable) charged particles (HSCPs) appear in
various extensions of the standard model
(SM)~\cite{Barbieri:2000vh,Appelquist:2000nn,Frampton:1997up,Dine:1994vc,Dine:1995ag,ArkaniHamed:2004fb,Feng:1999fu,Strassler:2006im}.   
If the lifetime of an HSCP produced at the Large Hadron Collider (LHC)
is longer than a few nanoseconds, the particle will travel over distances that
are comparable or larger than the size of a typical particle
detector. In addition,  if the HSCP mass is ${\gtrsim}100$ GeV/$c^2$, a
significant fraction of these particles will have a velocity, $\beta
\equiv v/c$, smaller than 0.9. These HSCPs will be directly observable
through the distinctive signature of a high  momentum ($p$) particle
with an anomalously large rate of energy loss through ionization ($dE/dx$). 

Previous collider searches for HSCPs have often been performed under the
assumption that these particles lose energy primarily through
low-momentum-transfer interactions, even if they are strongly
interacting, and are therefore likely to reach the
outer muon systems of the detectors and be identified as
muons~\cite{Aaltonen:2009kea, Abazov:2008qu,  Drees:1990yw,
Fairbairn:2006gg}.
The interactions with matter experienced by a strongly-interacting
HSCP, which is expected to form a bound state
($R$-hadron)~\cite{Farrar1978} in the process of hadronization, can
lead to it flipping the sign of its electric charge or becoming
neutral. A recent study~\cite{Mackeprang:2009ad} 
on the modeling of nuclear interactions of HSCPs traveling through
matter, favours a scenario of charge suppression. In this model the
probability is close to unity for an $R$-hadron containing a gluino,
\PSg\   (the supersymmetric partner of the gluon), 
or a supersymmetric bottom squark, to emerge as a neutral particle
after traversing an amount of material typical of the detectors
operating at LEP, the Tevatron, or LHC. If this model is correct, the
majority of these HSCPs would not be observed in the muon system of
a typical collider detector. Experimental strategies that do not rely
on the muon-like behavior for the HSCPs are therefore of great
importance. For instance, searches have been performed for very slow
($\beta \lesssim 0.4$) $R$-hadrons containing a gluino brought to rest in the 
detector~\cite{Abazov:2007ht, PhysRevLett.106.011801}. 

In this article we present a search for HSCPs 
produced in pp collisions at $\sqrt{s} = 7$ TeV at the LHC with the
Compact Muon Solenoid (CMS) detector~\cite{:2008zzk}. The search is based on
the data sample collected between April and August 2010 corresponding to an
integrated luminosity of $3.06$ pb$^{-1}$. 
We use triggers requiring: a high-transverse-momentum
muon ($p_T> 9$ GeV/$c$); or a dimuon pair ($p_T> 3$ GeV/$c$ for each muon); or
calorimeter-based missing transverse energy ($\MET >100$
GeV), to search for HSCPs failing muon identification or emerging
mainly as neutral particles after traversing the calorimeters;
or a high-transverse-energy jet ($E_T>100 $ GeV) to search for
HSCPs accompanied by substantial hadronic activity. 
The analysis makes use of two approaches. In a first selection,
referred to as ``\textit{tracker-only}'', the HSCP candidates are
searched for as individual tracks reconstructed in the inner tracker
detector with large $dE/dx$ and \pt. 
A second selection, referred to as ``\textit{tracker-plus-muon}'',   
additionally requires that the track is identified as a muon in the
outer muon detector. For both selections, the mass of the candidate is
calculated from the measured $p$ and $dE/dx$.

\section{The CMS Detector}
The central feature of the CMS detector is a 3.8 T superconducting
solenoid of 6 m internal diameter surrounding a silicon pixel and
strip tracker, 
a crystal electromagnetic calorimeter, and a
brass-scintillator hadronic calorimeter.  Muons are measured in
gaseous detectors embedded in the iron return yoke. 
Centrally produced charged particles are measured in the
tracker by three layers of silicon pixel detectors, followed by ten microstrip layers.
At pseudorapidities ($\eta \equiv -\ln \tan (\theta/2)$, where
$\theta$ is the polar angle measured with respect to the
beam direction) above $\approx
1.5$, particles are tracked in two pixel and twelve strip layers
arranged in disks perpendicular to the beam axis. In this
analysis, the $dE/dx$ measurement is based only on the
information from the silicon strip detectors. 
The $dE/dx$ measurement precision is limited by the silicon
strip analogue-to-digital converter (ADC) modules that are characterized by a
maximum number of counts per channel corresponding to about three
times the average charge released by a minimum--ionizing particle
(MIP) in 300 $\mu$m  of silicon. This is the thickness of the modules
mounted in the innermost silicon strip central layers.
The \pt\ resolution for tracks measured in the central (forward)
region of the silicon tracker is 1\% (2\%) for \pt\ values up to 50
GeV/$c$ and degrades to 10\% (20\%) at \pt\ values of 1 TeV/$c$.   
The trigger and reconstruction efficiencies for HSCPs in the muon
detectors are limited by the requirements on the arrival time of the
particles at the muon system. These requirements affect the efficiency
for detecting slow HSCPs. The dependence of the muon trigger
efficiency on the particle velocity ($\beta$) is studied using data
and Monte Carlo (MC) simulations and found to decrease, below
$\beta=0.7$. The muon trigger becomes completely inefficient at $\beta=0.5$.
A much more detailed description of the CMS apparatus can be found
elsewhere~\cite{:2008zzk}. 

\section{Candidate Selection and Background Estimation}
\begin{figure*}[htbp!]
\begin{center}
\includegraphics[width=0.45\textwidth]{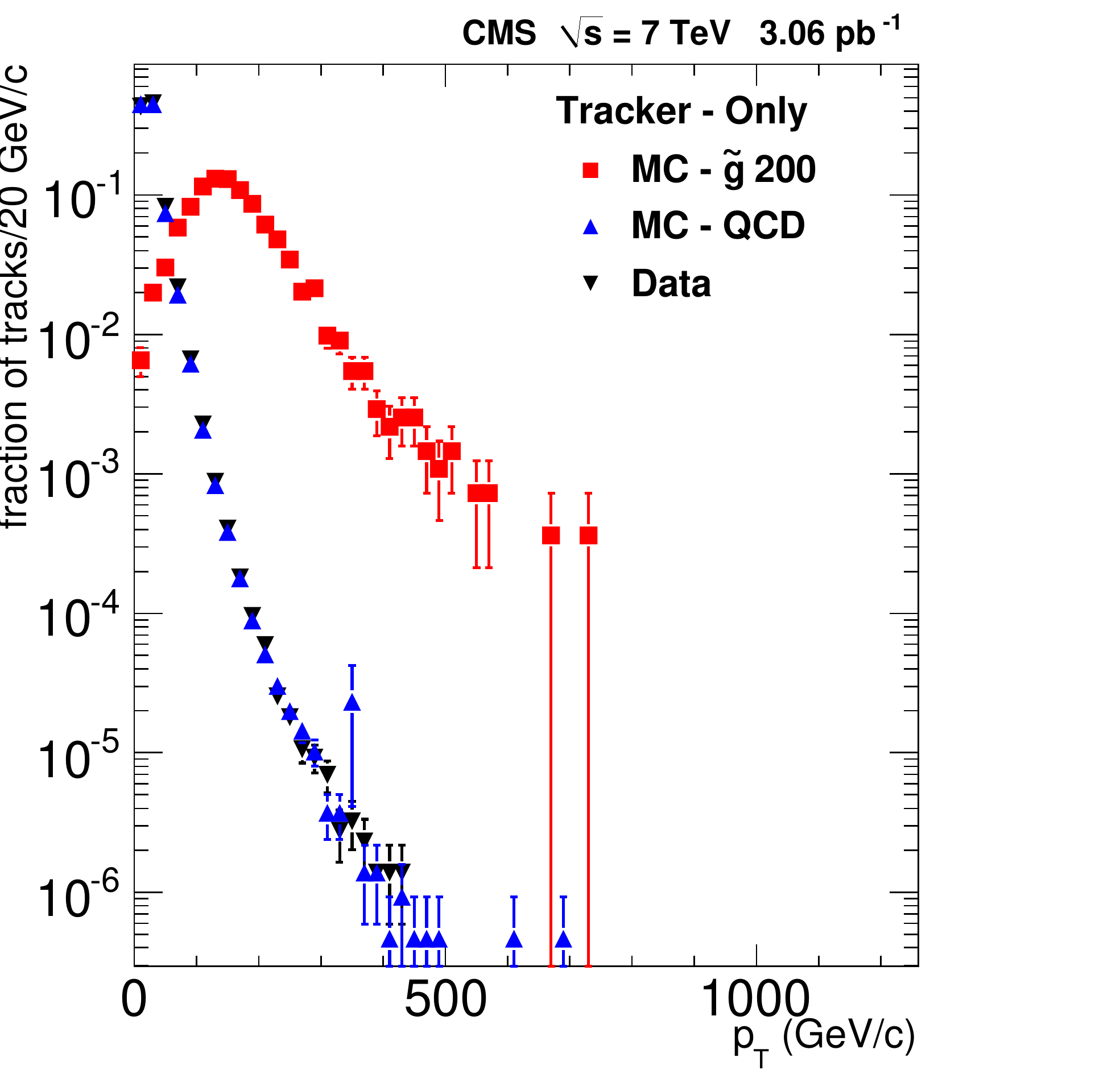}
\includegraphics[width=0.45\textwidth]{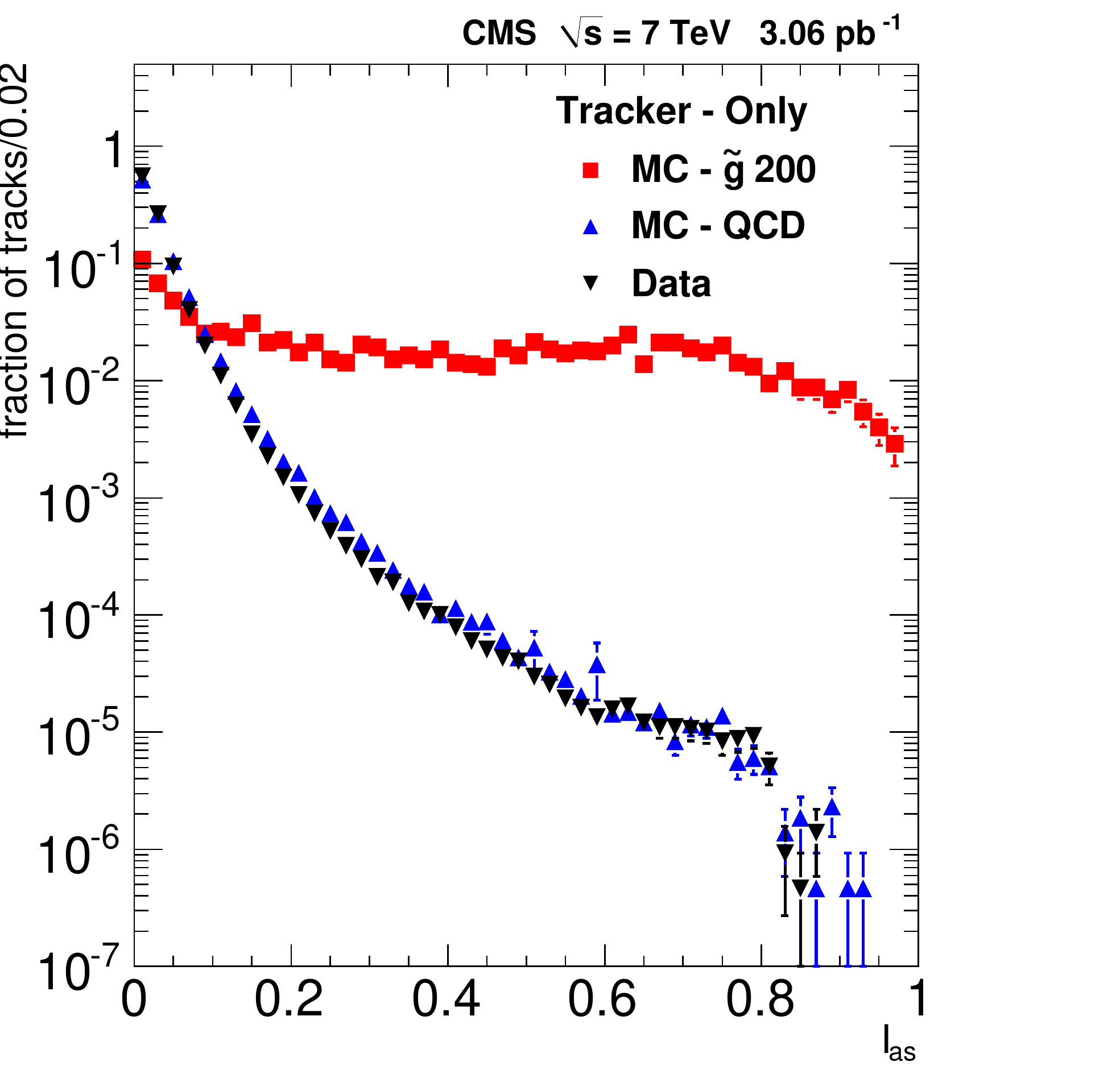}
 
\caption{Normalized distributions of \pt (left) and $I_{as}$ (right)
in data and two MC samples, for candidates passing the tracker-only
pre-selection. The two MC samples contain events from QCD processes
and from pair-production of \PSg\  with a mass of 200 GeV/$c^2$,
respectively.}  
\label{fig:datamc}
\end{center}
\end{figure*}

Candidate HSCPs are pre-selected by requiring a track with $|\eta| < 2.5$,
$p_T > 15$ GeV/$c$, relative uncertainty on the $p_T$ less than 15\%,
and transverse (longitudinal) impact parameter with respect to the
reconstructed primary collision vertex less than 0.25 (2.0)
cm. Candidate tracks are also required to have at least three measurements
in the silicon-strip detector. For the tracker-plus-muon selection, we
additionally require the track to be compatible
with track segments reconstructed in the muon system.
As an estimator of the degree of compatibility of the observed
charge measurements with the MIP hypothesis, 
a modified version of the Smirnov-Cramer-von Mises~\cite{Eadie, James}
discriminant is used (the modification applied to the original form of
the discriminant eliminates the sensitivity to incompatibility with the MIP
hypothesis due to low ionization):
\begin{equation}
I_{as} = \frac{3}{N} \times \left(
\frac{1}{12N} + \sum_{i=1}^N
\left[
P_i \times \left( P_i - \frac{2i-1}{2N} \right)^2 \right] \right),
\end{equation}
where $N$ is the number of charge measurements 
in the silicon-strip detectors, $P_i$ is the
probability for a MIP to produce a charge smaller or equal to the
$i$--th charge measurement for the observed path length in the detector, and the sum is over the track measurements ordered in terms of increasing $P_i$.
The charge probability density function used to calculate $P_i$ is
obtained using tracks with $p>5$ GeV/$c$ in events collected with a
minimum bias trigger. Non-relativistic HSCP candidates will have the
value of the discriminant $I_{as}$ approaching unity.  
Figure~\ref{fig:datamc} shows normalized distributions of \pt and
$I_{as}$ in data and two MC samples, for candidates passing the
tracker-only pre-selection. The first MC sample contains events
from QCD processes. The second MC sample contains signal events
from pair-production of stable \PSg\  with a mass of 200 GeV/$c^2$. Both
samples are generated with the \PYTHIA
v6.422~\cite{Sjostrand:2006za} MC package. More details on the
simulation of the signal sample will be given below. 
The MC QCD simulations are found to reproduce the data, and the
simulated signal is clearly separated.
Because of the limited number of available simulated events with low 
transverse-momentum transfers, the  MC QCD distributions display 
bin-to-bin variations in the size of the statistical errors.  

The most probable value of the particle $dE/dx$ is determined using a
harmonic estimator $I_h$ of grade $k=-2$:    
\begin{equation}
I_h= \biggl( \cfrac{1}{N} \sum_i c_{i}^{k} \biggr)^{1/k},
\label{eq:HarmonicEstimator}
\end{equation} 
where $c_{i}$ is the charge per unit path length in the detector 
of the $i$-th measurement for a given track.
In order to estimate the mass ($m$) of highly ionizing particles,  
the following relationship between $I_h$, $p$,
and $m$ is assumed: 
\begin{equation}
I_h= K\cfrac{m^2}{p^2}+C.
\label{eq:MassFromHarmonicEstimator}
\end{equation}
Equation~\ref{eq:MassFromHarmonicEstimator} reproduces the Bethe-Bloch
formula~\cite{Nakamura:2010zzi} with an accuracy of better than 1\% in
the range $0.4 < \beta < 0.9$, which corresponds to $1.1 <
(dE/dx)/(dE/dx)_{MIP} < 4.0$. The empirical parameters $K$ and $C$ are
determined from data using a sample of low-momentum protons, for which
the fitted values are $K=2.579 \pm 0.001$ MeV cm$^{-1}$ $c^2$ and
$C=2.557 \pm 0.001$ MeV cm$^{-1}$, and the mass resolution is 7\%.
The reconstructed mass distribution for kaons and protons is in
very good agreement with the one obtained from MC following this
procedure~\cite{vertex}.
For masses above 100 GeV/$c^2$, the mass resolution is expected to
worsen  because of the deterioration of the momentum
resolution and because of the limit on the maximum charge that can be
measured by the silicon strip tracker ADCs, which also affects the mass scale. 
For a 300 GeV/c$^2$ HSCP, the  mass resolution is
12\% and the reconstructed peak position is at 265 GeV/$c^2$. 

The search is performed as a counting experiment. 
Signal candidates are required to have $I_{as}$ and
$p_T$ greater than threshold values and the
mass to be in the range of 75 to 2000 GeV/$c^2$, allowing sensitivity to HSCP
masses as low as 100 \GeVcc.
The $I_{as}$ distribution for the pre-selected tracks, and in particular its
tail, depends strongly on the number of charge measurements on the track. 
Thus, to increase the sensitivity of the search, pre-selected tracks
are divided into subsamples according to the number of silicon strip
measurements.
The $I_{as}$ ($p_T$) threshold in each subsample is determined by
requiring a constant efficiency on data for all subsamples, when the
threshold is applied separately. 
A method that exploits the absence of correlation  between the 
$p_T$ and $dE/dx$ measurements in data is used to estimate the background
from MIPs. In a given subsample $j$, the number of tracks that are
expected to pass both
the final $p_T$ and $I_{as}$ thresholds set for the subsample is
estimated as $D_j=B_jC_j/A_j$, where $A_j$ is the number of tracks
that fail both the $I_{as}$ and $p_T$ selections and $B_j$
($C_j$) is the number of tracks that pass only the $I_{as}$ ($p_T$)
selection.
The $B_j$ and $C_j$ tracks are then used to form a binned 
probability density function in $I_h$ ($p$) for the $D_j$ tracks. 
Finally, using the mass determination
(Eq.~\ref{eq:MassFromHarmonicEstimator}), the full
mass spectrum of the background in the signal region $D$ is
predicted. 

By comparing the predicted and observed number of tracks for
several very loose selections in a control region of the mass spectrum,
corresponding to masses below 75 GeV/$c^2$, the prediction is found to 
underestimate systematically the observation by 12\% (5\%) for the
tracker-only (tracker-plus-muon) selection. 
After correcting the predicted background by this amount, the remaining
background systematic uncertainty is conservatively estimated as twice the
r.m.s. of the prediction-to-observation ratio distribution
The resulting uncertainty on the predicted background is 14\% (17\%).

As significant background rejection can be obtained without a sizable
effect on the signal efficiency, the final selection is optimized by
requiring the total expected background in the search region 
to be $\sim 0.05$ events. This low-background choice optimizes the 
discovery potential even if just a handful of events are
observed, and at the same time maintains significant exclusion
sensitivity in the case that no events are observed.

\section{Results}
\begin{figure*}[htbp]
\begin{center}
\includegraphics[width=0.45\textwidth]{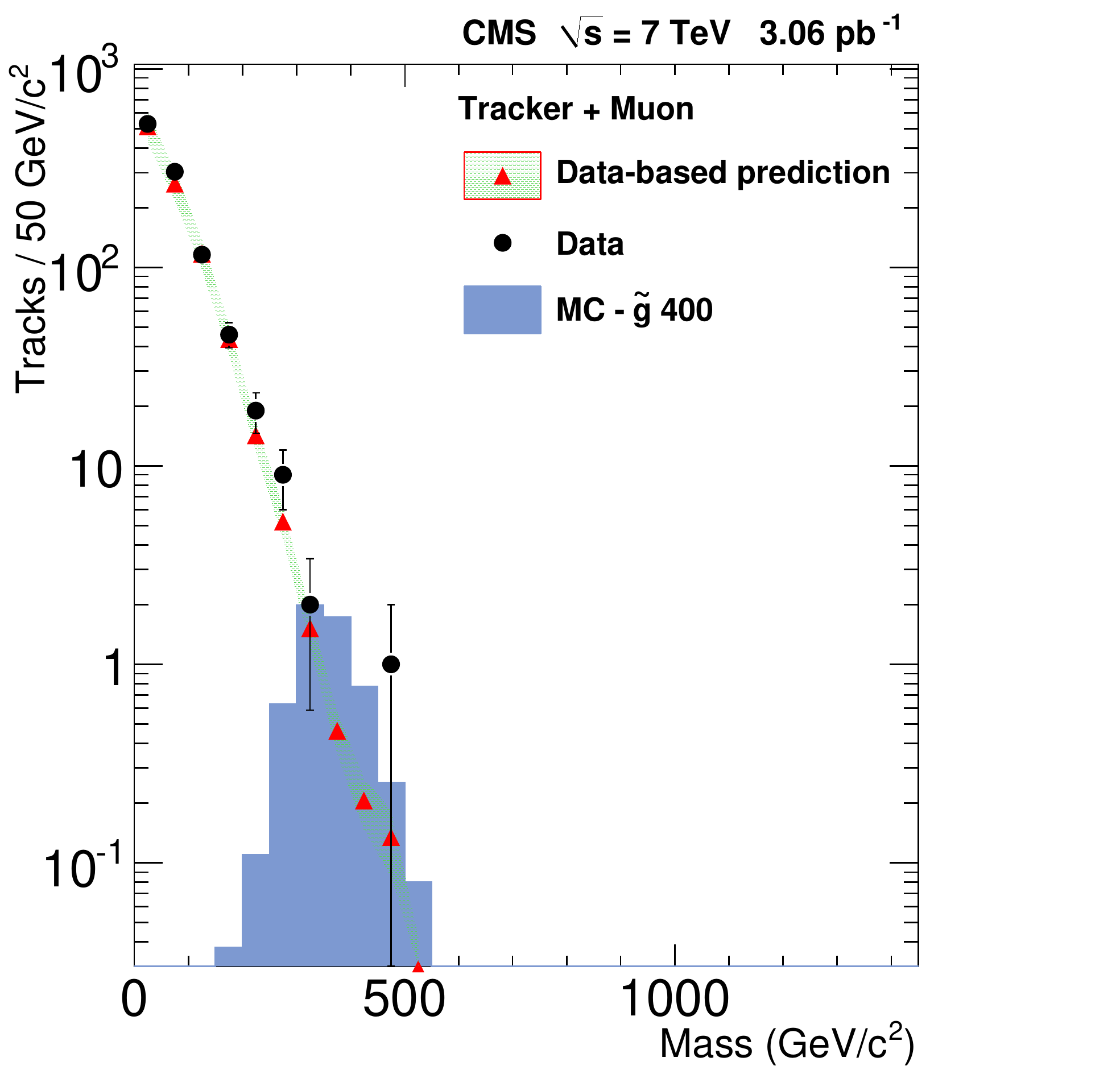}
\includegraphics[width=0.45\textwidth]{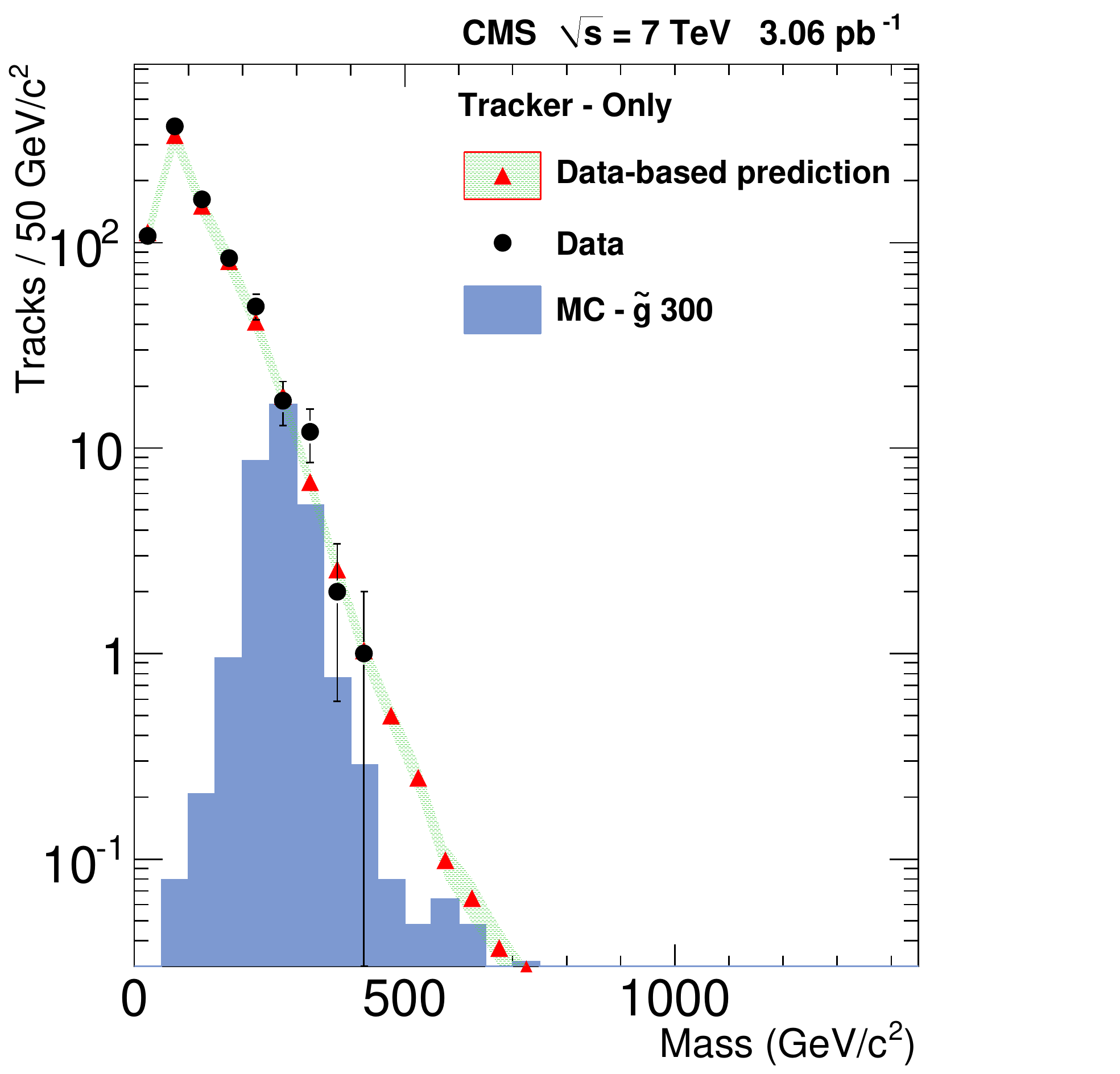}
\caption{Mass spectrum for the loose selection defined in
Table~\ref{tab:selectioneff} for the tracker-plus-muon (left) and tracker-only
(right) candidates.
Shown are: observed spectrum (black dots with the error bars), data-based
predicted background spectrum (red triangles) with its uncertainty
(green band) and the spectrum predicted by MC for a signal
of pair-produced stable \PSg\  with a mass of 400 (left) and
300 (right) GeV/$c^2$ (blue histogram).
}
\label{fig:ts_bgdshapeprediction}
\end{center}
\end{figure*}

In addition to the final ``\textit{tight}'' selection, the result of a
``\textit{loose}'' selection is reported in
Table~\ref{tab:selectioneff}. The loose  selection retains a
relatively large number of background candidates and allows us to
compare the background prediction with the observed data.
Figure~\ref{fig:ts_bgdshapeprediction} shows good agreement between
the observed and predicted mass spectrum obtained using the loose
selection for the tracker-plus-muon and tracker-only candidates. 

\begin{table}
\begin{center}
\caption{\label{tab:selectioneff} Selections used in the
analysis and results of the search. The tracker-plus-muon and
tracker-only selections are labeled as ``Mu'' and ``Tk'',
respectively.  As explained in the text, the
actual $I_{as}$ (\pt) thresholds are determined in the various
subsamples by the requirement of a constant efficiency for candidate
selection, $\epsilon_{I}$ ($\epsilon_{p_T}$).  
These thresholds, indicated by
$I_{as}^{min}$ ($p_T^{min}$), are therefore reported as a range of
values. Expected and observed number of candidates in the signal
region are reported in the ``Expected'' and ``Observed'' rows,
respectively. 
Top: loose selection. Bottom: tight selection.}

\begin{tabular}{| l | c | c | } \hline
{\bf LOOSE}                 &  \bf Mu                   & \bf Tk    \\ \hline
$\epsilon_{I}$        &  $3.2 \times 10^{-2}$ & $1.0 \times 10^{-2}$
\\ \hline
$I_{as}^{min} $       &  0.049 - 0.162        &  0.007 - 0.278  \\ \hline 
$\epsilon_{p_T}$      &  $1.0 \times 10^{-1}$ & $3.2 \times 10^{-2}$  \\ \hline
$p_T^{min}$ (GeV/$c$) &  34 -  36             &  59 -  62 \\ \hline
Expected & $281 \pm  2 (stat.) \pm 49 (syst.) $     & $426 \pm  1 (stat.) \pm 62 (syst.) $   \\ \hline
Observed  & 307   &  452 \\ 
\hline
\hline
{\bf TIGHT}                 &  \bf Mu                   & \bf Tk    \\ \hline
$\epsilon_{I}$        &  $1.0 \times 10^{-4}$ & $1.0 \times 10^{-4}$ \\ \hline
$I_{as}^{min} $       &  0.184 - 0.782         &  0.186 - 0.784   \\ \hline 
$\epsilon_{p_T}$      &  $1.0 \times 10^{-3}$ & $3.2 \times 10^{-4}$  \\ \hline
$p_T^{min}$ (GeV/$c$) &  115 -  118             &  154 -  210 \\ \hline
Expected &  $0.025 \pm 0.002 (stat.) \pm 0.004 (syst.) $    & $0.074 \pm 0.002 (stat.) \pm 0.011 (syst.) $  \\ \hline
Observed  &  0    & 0 \\ \hline
\end{tabular}
\end{center}
\end{table}

The results of the search with the final selection are also presented in 
Table~\ref{tab:selectioneff}. No candidate HSCP track
is observed in either the tracker-only or tracker-plus-muon analysis. 

Given the null result, cross section upper limits at the 95\%
C.L. are set on the HSCP production for two benchmark
scenarios: direct production of \PSg\ pairs and \stone pairs.
For a given mass,  the cross section for \PSg\ production is expected
to be much larger than that for \stone production at both the Tevatron
and the LHC.  Thus higher mass limits can be set for the former at
both machines. However, as the mass of a produced particle increases,
the ratio of the production cross section at the LHC to that
at the Tevatron increases. For \PSg\ masses in the region of 350
GeV/$c^2$, the increase in relative cross section outweighs the difference
in integrated luminosity between the current Tevatron and LHC data
sets, enabling the LHC to set the most sensitive limits on the search
for \PSg.

Events with pair production of \PSg\ and \stone, with mass values in the
range 130-900 GeV/$c^2$, are generated with \PYTHIA in order to
compute the efficiency of our selection on these signals. 
The \stone\ and \PSg\  are treated as stable in all these
samples and their hadronization is performed by \PYTHIA. 
A parameter relevant to the \PSg\  pair production, and not to the
\stone pair production, is  the fraction, $f$, of produced \PSg\   hadronizing
into a \PSg-gluon state ($R$-gluonball). This fraction is an unknown
parameter of the hadronization model and affects the fraction of
$R$-hadrons that are neutral at production, which in turn affects the
detection efficiency.  In this study, results are obtained for two
different values of $f$, 0.1 and 0.5, to show the effect of the
hadronization model uncertainty on the sensitivity of the search.
The interactions of the HSCPs with the CMS apparatus and the detector
response are simulated in detail with the \GEANTfour
v9.2~\cite{Agostinelli2003250, geant} toolkit. 
The $R$-hadron strong interactions with matter are modeled as in
Ref.~\cite{Kraan:2004tz, Mackeprang:2006gx}. 
This model, like a number of others~\cite{deBoer:2007ii, Baer:1998pg,
Mafi:1999dg, Mackeprang:2009ad}, assumes that the 
probability of an interaction between the heavy
parton and a quark in the target nucleon is low since the
cross section varies with the inverse square of the parton
mass according to perturbative QCD. 
The adopted model chooses a pragmatic approach based on analogy with
observed low energy hadron scattering. 
However, given the very large uncertainties
on the dynamics underlying $R$-hadron interactions, an extremely
pessimistic scenario of complete charge suppression, where each nuclear
interaction suffered by the $R$-hadron causes it to become neutral, is
also considered.  The tracker-only selection is expected to have
sensitivity even in such a scenario. 
The total signal efficiency is reported in
Table~\ref{tab::stopDataTkMuon} for some
combinations of models and selections. 
\begin{table}
\begin{center}
\caption{\label{tab::stopDataTkMuon} Total signal selection efficiency
and cross section upper limits for different combinations of
models and  selections: pair production of supersymmetric stop and gluinos; tracker-plus-muon (Mu) and tracker-only (Tk) selections;
different fractions, $f$, of $R$-gluonball states produced after
hadronization and charge suppression (ch. suppr.) scenario.}
\begin{tabular}{| l | c c c c c c | } \hline
{\bf gluino} mass (GeV/$c^2$)         & 200 & 300 & 400 & 500 & 600 & 900 \\ \hline
Theoretical cross section (pb)  & 606 & 57.2 & 8.98 & 1.87 & 0.470 & 0.0130 \\
\hline
Mu; f=0.1  & & & & & & \\
Total efficiency (\%)         & 7.17 & 10.4 & 13.1 & 15.1 & 14.5 & 9.18 \\
Expected 95\% C.L. limit (pb) & 15.1 & 10.4 & 8.25 & 7.16 & 7.47 & 11.8 \\
Observed 95\% C.L. limit (pb) & 14.5 & 9.98 & 7.92 & 6.88 & 7.17 & 11.3 \\
\hline
Mu; f=0.5;  & & & & & & \\
Total efficiency (\%)         & 3.84 & 5.46 & 7.03 & 8.23 & 8.10 & 4.98 \\
Expected 95\% C.L. limit (pb) & 28.2 & 19.8 & 15.4 & 13.1 & 13.3 & 21.7 \\
Observed 95\% C.L. limit (pb) & 27.1 & 19.0 & 14.8 & 12.6 & 12.8 & 20.9 \\
\hline
Tk; f=0.1; ch. suppr.  & & & & & & \\
Total efficiency (\%)         & 0.59 & 2.44 & 4.16 & 6.39 & 8.60 & 7.66 \\
Expected 95\% C.L. limit (pb) & 188 & 45.5 & 26.7 & 17.4 & 12.9 & 14.5 \\
Observed 95\% C.L. limit (pb) & 176 & 42.6 & 25.0 & 16.2 & 12.1 & 13.6 \\
\hline
\hline
{\bf stop} mass (GeV/$c^2$)           & 130 & 200 & 300 & 500 & 800 &        \\ \hline
Theoretical cross section (pb)  & 120 & 13.0 & 1.31 & 0.0480 & 0.00110
&       \\ \hline
Mu;  & & & & & & \\
Total efficiency (\%)         & 2.99 & 9.50 & 14.7 & 19.6 & 14.0 &       \\
Expected 95\% C.L. limit (pb) & 36.1 & 11.4 & 7.35 & 5.52 & 7.71 &       \\
Observed 95\% C.L. limit (pb) & 34.7 & 10.9 & 7.06 & 5.30 & 7.39 &       \\
\hline
Tk; ch. suppr.  & & & & & & \\
Total efficiency (\%)         & 0.02 & 1.19 & 3.55 & 7.27 & 7.68 &       \\
Expected 95\% C.L. limit (pb) & 5540 & 93.2 & 31.3 & 15.3 & 14.5 &       \\
Observed 95\% C.L. limit (pb) & 5180 & 87.2 & 29.2 & 14.3 & 13.5 &       \\
\hline
\end{tabular}
\end{center}
\end{table}
Relatively small differences are found between the tracker-plus-muon
and tracker-only selection except in the charge suppression scenario,
where the tracker-plus-muon selection is completely inefficient.

This analysis is found to be complementary to the search for
long-lived stopped particles presented
in~\cite{PhysRevLett.106.011801}. Indeed, 
for the case of \PSg\  with $f=0.1$ and mass values below 500
GeV/$c^2$, the fraction of HSCPs that have $\beta<0.4$ and pass
the final selection is less than 0.5\%.  
Therefore the two analyses explore different ranges of
produced particle velocities with no overlap.

The main sources of systematic uncertainty affecting the
results presented in the following are summarized in
Table~\ref{tab::systematicerror}. 
\begin{table}
\begin{center}
\caption{\label{tab::systematicerror} Sources of systematic errors
and corresponding relative uncertainties.} 
\begin{tabular}{| l | c | } \hline
Source of Systematic Error          & Relative Uncertainty ($\%$)  \\
\hline \hline
Theoretical cross section & 10 - 25 \\ \hline
Integrated luminosity & 11  \\ \hline \hline
Trigger efficiency & $12$ \\
Muon reconstruction efficiency & 5 \\
Track reconstruction efficiency & $<5$\\
Momentum scale & $< 5$ \\
Ionization energy loss scale & $< 3$ \\
\hline
Total uncertainty on signal acceptance & 15 \\ \hline
 \end{tabular}
\end{center}
\end{table}
The uncertainty on the signal selection efficiency is estimated to be
15\% for all considered combinations of models and selections. The
main source of this uncertainty is an assumed 10\%
uncertainty on the jet energy scale~\cite{JME-10-003}, which affects
both the jet and $E_T^{miss}$ trigger efficiency by about 10\%. In a
more recent study~\cite{JME-10-010}, the estimate of the uncertainty
on the jet energy scale has been reduced by a factor of two. However,
in this analysis we have conservatively chosen to retain the earlier
estimate of 10\%.  
The uncertainty on the muon trigger efficiency and the impact of an
imperfect simulation of the synchronization of the muon trigger and
readout electronics are studied with data and MC. They result in an
overall uncertainty on the signal selection efficiency of less than 5\%.  
The uncertainty on the offline track reconstruction
efficiency~\cite{TRK-10-002, MUO-10-002}, track momentum
scale~\cite{TRK-10-004}  and ionization energy loss scale~\cite{vertex} 
is also found to yield no more than 5\%
uncertainty on the overall signal selection efficiency. The uncertainty on the
absolute value of the integrated luminosity is estimated to be 11\%
~\cite{EWK-10-004}.

\begin{figure*}
\begin{center}
\includegraphics[width=0.95\textwidth]{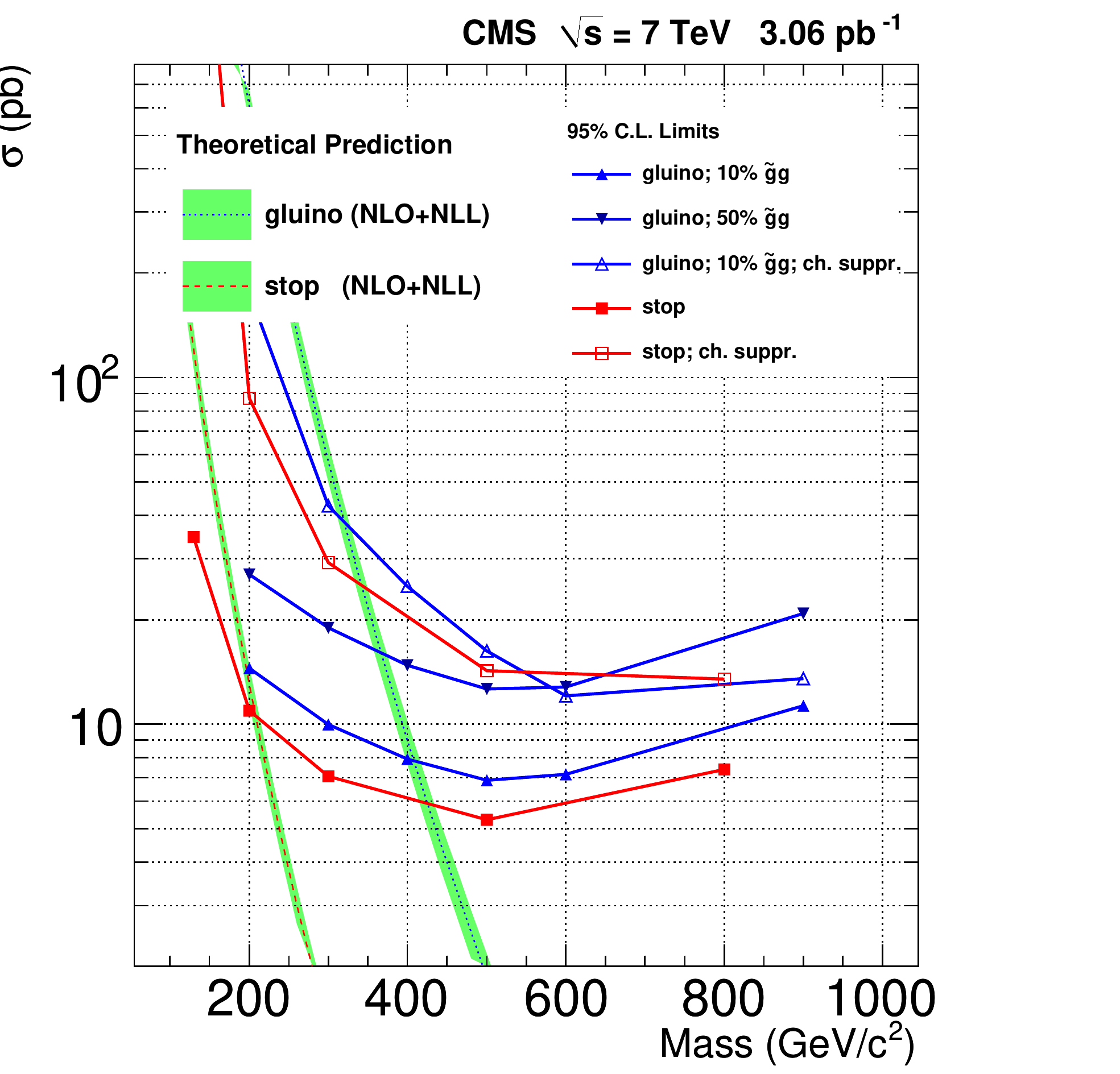}
\caption{\label{fig::STOPMassExclusion} 
Predicted theoretical cross section and observed 95\% C.L. upper limits on
the cross section for the different combinations of models
and scenarios considered: pair production of supersymmetric stop and gluinos;
different fractions, $f$, of $R$-gluonball states produced after hadronization
and charge suppression (``ch. suppr.'') scenarios. Only the results
obtained with the most sensitive selection are reported: tracker-only
for the charge suppression scenarios and tracker-plus-muon
for all other cases.
The bands represent the theoretical 
uncertainties on the cross section values.  
}
\end{center}
\end{figure*}

The upper limit on the cross section is computed at 95\% C.L. using a
Bayesian method with a flat signal prior and a log-normal
prior used for integration over the nuisance parameters~\cite{Nakamura:2010zzi,Eadie, James}.
In order to obtain a conservative upper limit we set the expected
background to zero.
The tracker-plus-muon selection provides better limits than the
tracker-only for all scenarios but the one with complete charge
suppression. For each considered scenario, the cross section
upper limit obtained with the most sensitive selection 
is reported in Table~\ref{tab::stopDataTkMuon} and 
Fig.~\ref{fig::STOPMassExclusion},
along with the theoretical predictions for \PSg\  and \stone pair
production computed at next-to-leading order (NLO) + next-to-leading
log (NLL)~\cite{Kulesza:2008jb,Kulesza:2009kq, 
Beenakker:2009ha, Beenakker:2010nq} using
the \PROSPINO v2 program~\cite{Beenakker:1996ed}.
The \PSg\  theoretical predictions refer to the case where the squarks
and gluino are degenerate in mass. In the heavy squark limit these
cross sections are about 10\% higher.  
For the case of \stone, beyond LO, the cross section does not only depend
on the \stone mass, but also, though to a much lesser
extent~\cite{Beenakker:1997ut}, on the \PSg\  mass, the average mass
of the first and second generation squarks and the stop mixing angle.
For this reason, the \stone theoretical predictions reported in
Table~\ref{tab::stopDataTkMuon} and Fig.~\ref{fig::STOPMassExclusion}
refer to the SPS1a\textquoteright \ benchmark
scenario~\cite{AguilarSaavedra:2005pw}.   
All systematic uncertainties discussed above are included in the cross
section upper limits reported in Table~\ref{tab::stopDataTkMuon} and
Fig.~\ref{fig::STOPMassExclusion}. 
From the intersection of the cross section limit curve  
and the lower edge of the theoretical cross section band we set a 95\% C.L.
lower limit of 398 (357) GeV/$c^2$ on the mass of
pair-produced \PSg\  with $f=0.1 (0.5)$, using the tracker-plus-muon
selection.
The analogous limit on the \stone\ mass is 202 GeV/$c^2$.
In the charge suppression scenario we set, with the tracker-only
selection, a 95\% C.L. \PSg\  mass limit of 311 GeV/$c^2$ for $f=0.1$. 

\section{Conclusions}
In summary, the CMS detector has been used
to identify highly ionizing, high-$p_T$ particles and measure
their masses. Two searches have been conducted: a very inclusive and
model independent one that uses highly-ionizing tracks reconstructed
in the inner tracker detector, and another requiring also that these tracks
be identified in the CMS muon system.
In each case, the observed distribution of the candidate masses is
consistent with the expected background. We have set
lower limits on masses of stable strongly interacting supersymmetric 
particles.
For the case of \PSg\  with $f=0.1$ and \stone, a lower
mass limit of 398 and 202 GeV/$c^2$, respectively, is set at the
95\% C.L. with the analysis that uses muon identification. 
In a pessimistic scenario of complete charge suppression the above
\PSg\  mass limit is reduced to 311 GeV/$c^2$ and is obtained with
the tracker-only selection.  
The limits presented here on stable \PSg\ are the most restrictive to
date.

\section{Acknowledgements}
We are grateful to Anna Kulesza and Michael Krammer for providing the
theoretical production cross sections and associated uncertainties
at next-to-leading order for pair production of \PSg\  and \stone . 
We wish to congratulate our colleagues in the CERN accelerator departments for the excellent performance of the LHC machine. We thank the technical and administrative staff at CERN and other CMS institutes, and acknowledge support from: FMSR (Austria); FNRS and FWO (Belgium); CNPq, CAPES, FAPERJ, and FAPESP (Brazil); MES (Bulgaria); CERN; CAS, MoST, and NSFC (China); COLCIENCIAS (Colombia); MSES (Croatia); RPF (Cyprus); Academy of Sciences and NICPB (Estonia); Academy of Finland, ME, and HIP (Finland); CEA and CNRS/IN2P3 (France); BMBF, DFG, and HGF (Germany); GSRT (Greece); OTKA and NKTH (Hungary); DAE and DST (India); IPM (Iran); SFI (Ireland); INFN (Italy); NRF and WCU (Korea); LAS (Lithuania); CINVESTAV, CONACYT, SEP, and UASLP-FAI (Mexico); PAEC (Pakistan); SCSR (Poland); FCT (Portugal); JINR (Armenia, Belarus, Georgia, Ukraine, Uzbekistan); MST and MAE (Russia); MSTD (Serbia); MICINN and CPAN (Spain); Swiss Funding Agencies (Switzerland); NSC (Taipei); TUBITAK and TAEK (Turkey); STFC (United Kingdom); DOE and NSF (USA).

\bibliography{auto_generated}   

\cleardoublepage\appendix\section{The CMS Collaboration \label{app:collab}}\begin{sloppypar}\hyphenpenalty=5000\widowpenalty=500\clubpenalty=5000\textbf{Yerevan Physics Institute,  Yerevan,  Armenia}\\*[0pt]
V.~Khachatryan, A.M.~Sirunyan, A.~Tumasyan
\vskip\cmsinstskip
\textbf{Institut f\"{u}r Hochenergiephysik der OeAW,  Wien,  Austria}\\*[0pt]
W.~Adam, T.~Bergauer, M.~Dragicevic, J.~Er\"{o}, C.~Fabjan, M.~Friedl, R.~Fr\"{u}hwirth, V.M.~Ghete, J.~Hammer\cmsAuthorMark{1}, S.~H\"{a}nsel, C.~Hartl, M.~Hoch, N.~H\"{o}rmann, J.~Hrubec, M.~Jeitler, G.~Kasieczka, W.~Kiesenhofer, M.~Krammer, D.~Liko, I.~Mikulec, M.~Pernicka, H.~Rohringer, R.~Sch\"{o}fbeck, J.~Strauss, A.~Taurok, F.~Teischinger, W.~Waltenberger, G.~Walzel, E.~Widl, C.-E.~Wulz
% ...
\vskip\cmsinstskip
% ...
1:~~Also at CERN, European Organization for Nuclear Research, Geneva, Switzerland\\
% ...

\end{sloppypar}
\end{document}